# Computational inverse design for cascaded systems of metasurface optics

ADAM S. BACKER[1,*]

[1] *Sandia National Laboratories, New Mexico, P.O. Box 5800, Albuquerque, NM 87185-1413, USA*
*\*abacke@sandia.gov*

**Abstract:** Metasurfaces are an emerging technology that may supplant many of the conventional optics found in imaging devices, displays, and precision scientific instruments. Here, we develop a method for designing optical systems composed of multiple unique metasurfaces aligned in sequence. Our approach is based on computational inverse design, also known as the adjoint-gradient method. This technique enables thousands or millions of independent design variables to be optimized in parallel, with little or no intervention required by the user. To demonstrate the broad applicability of our method, we use it to design an achromatic doublet metasurface lens, a spectrally-multiplexed holographic element, and an ultra-compact optical neural network for classifying handwritten digits.

## 1. Introduction

Optical metasurfaces are composite structures that utilize nanoscale patterning to achieve properties not found in nature [1,2]. By tailoring the dimensions of nanoscale scattering elements (meta-atoms) arranged in a periodic lattice, it is possible to precisely shape the wavefronts (phase) of incident light (see Fig. 1(a) for a representative image of a metasurface). Using metasurfaces based on dielectric meta-atoms, high numerical aperture lenses have been demonstrated with high transmission and focusing efficiency, as well as holograms, beam steerers, and gratings [3-8]. While glass optics bend light using macroscale curvature, metasurfaces feature a nearly-flat form factor, and are not subject to the same geometric design constraints. Metasurfaces have the potential to streamline complex optical assemblies, reducing their bulkiness, weight, and cost, and may be used to achieve fundamentally novel functionalities that cannot be realized by conventional means [9-11].

The design of high-performance metasurface optics is an open problem that spans the fields of computational electromagnetics, optical engineering, and numerical optimization. While single-wavelength metasurfaces may be specified using ad-hoc approximations and parameter sweeps over a limited set of meta-atom geometries, the design of polychromatic and multi-functional devices is not as clear-cut. To address these issues, we propose a method for designing optical systems composed of multiple unique metasurfaces cascaded along the optical propagation axis [12,13]. Our method utilizes simple, polarization-independent meta-atoms as the basic building-blocks for our design. Yet, we discover capabilities that would not be possible using "singlet" metasurfaces composed of just one diffracting element (and the same library of meta-atom geometries). Our approach relies on a powerful optimization technique called computational inverse design [14,15], also known as the adjoint gradient method (furthermore, in the field of machine learning, the method is similar to that of *backpropagation*, which is often used for training neural networks). Inverse design provides a means of calculating the gradients of a cost function with respect to *all* design variables by solving a so-called "adjoint problem." The adjoint problem has similar computational complexity to a single solution of the "forward problem," which is used to evaluate the cost function at a given design iteration. The advantage of inverse design is the fact that the computational complexity of evaluating the gradients depends only on that of the forward problem, even as the number of design variables increases. Hence, many parameters may be optimized in parallel. Computational inverse design has found applications within a multitude of engineering disciplines [16,17]. The technique has been used to design electromagnetic and optical devices, and revolutionary silicon photonics components [18-28]. Recently, researchers have recognized the utility of inverse-design-based approaches for metasurface optics [29-31], among other methods [32]. Such work has led to designs for polychromatic and high-incidence-angle metasurface lenses and gratings [29, 33-35], and has facilitated the exploration of novel fabrication platforms, such as arrangements of Mie-scattering dielectric spheres [36]. Notably, recent works have considered the optimization of metasurfaces composed of multilayer meta-atoms [37-39]. In such designs, differently-shaped dielectric scatterers are stacked on top of each other within a subwavelength volume, in order to achieve superior performance to single-layer meta-atoms. In contrast, this paper considers only simple meta-atom geometries (rectangular titanium dioxide ($TiO_2$) nanopillars of different widths), and instead leverages the interplay between multiple metasurfaces spaced distances $\gg \lambda$ apart to realize capabilities not possible using singlet metasurfaces. This work should be viewed as complementary to previous approaches, in that the method described here may be applied to the large-scale optimization of an entire optical system containing multiple metasurface elements, while alternative, nanoscale topology-optimization methods may concurrently be used to significantly improve the geometries of the individual meta-atoms that compose a particular metasurface.

Our inverse design framework relies heavily upon Fourier optics, and the underlying mathematics are very similar to the analytical gradient calculations used in some phase-retrieval algorithms [40-42]. Here we are primarily interested in designing novel optical systems, as opposed to characterizing optical aberrations from a diverse set of images. Secondly, we demonstrate how the optimization framework may be extended to enable computationally-efficient Fourier optics simulations to be combined with finite-difference-time-domain (FDTD) parameter sweeps to directly optimize a complex optical system with respect to individual meta-atom geometries. Finally, although this paper is primarily concerned with metasurface design, the methods may be applied to optical systems containing multiple air-spaced diffractive optics, photolithographically fabricated phase masks, or spatial light modulators.

This paper is organized as follows: In Section 2 we briefly summarize other approaches to metasurface design and introduce a distinction between what we term the *macroscale* and *nanoscale* design problems. In Section 3 we develop the computational inverse design framework. In Section 4 we apply our method to a range of applications: We design an achromatic doublet lens for use in the visible spectrum, a spectrally multiplexed holographic element, and a compact optical neural network for classifying handwritten digits. In Section 5, we discuss future extensions and generalizations of the method.

## 2. Relation to previous work

A key difficulty encountered in metasurface design problems is their inherently multiscale nature: It is desirable to build devices with macroscale dimensions (from 100s of μm up to cm in size), but it is also necessary to precisely tailor the geometries of nanoscale structures (individual meta-atoms) in order to achieve the intended function of the macroscale device. Rigorously modeling an entire metasurface, using e. g. finite difference time domain (FDTD) methods, simultaneously over the nanoscale and macroscale length-scales is computationally prohibitive, due to the narrow simulation mesh sizes required, and formidable memory requirements. To address this issue, it is often customary to divide the task of designing a metasurface into what we term a *macroscale* problem and a *nanoscale* problem. In the macroscale problem, some desired output to the overall metasurface device specified, usually in the form of a desired wavefront or phase function. For example, in the case of a singlet metasurface lens, a spherically converging phase would be specified, or in the case of a beam deflector, a linear phase ramp would be required. In the nanoscale problem, the individual meta-atom geometries must be chosen to realize the desired output at each spatial location throughout the metasurface. This step is normally performed by simulating a library of meta-atom geometries (using e. g. an FDTD parameter sweep), and recording output parameters of interest such as phase and transmission into a lookup-table [5,43-46]. Since each simulation used to construct the lookup-table is performed over the dimensions of just a single meta-atom, a complete parameter-sweep over every design variable is computationally tractable. A prescription for a complete metasurface optic can subsequently be specified by choosing meta-atoms from the lookup table that most closely match the desired output at all lattice positions across the metasurface aperture. This design strategy assumes that there is minimal interaction between adjacent meta-atoms, or the metasurface structure is locally periodic (meta-atom geometry does not change drastically over small sections of a given metasurface). Other approaches for metasurface design have involved the use of the Pancharatnam-Berry (geometric) phase [6,7,47-49], asymmetric polarization-dependent meta-atom geometries [6,50-52], Huygens' surfaces [53-55], and libraries of complex meta-atom shapes used to achieve unique dispersion properties [56]. For simplicity, in this paper, we focus on simple, polarization-independent meta-atoms (square-shaped nanostructures of varying widths). However, in Section 5, we discuss how our method could be potentially extended to consider more complex meta-atom shapes, possibly improving overall design performance.

The above described approaches to metasurface design have been used in many contexts and are highly effective when the solution to the macroscale problem is intuitive and can be easily specified. This is the case when designing singlet metasurface lenses, since the phase function associated with an ideal thin lens is already known. Challenges arise, however, when the solution to the macroscale problem is not obvious. If an optical system is composed of multiple cascaded metasurfaces, it is not clear what metasurface phase should be specified at each layer of the optical system to achieve ideal performance. This problem becomes especially difficult when polychromatic functionality is desired – in this case, different desired phases must be selected for different wavelengths.

Recently, methods for designing two-layer, single-wavelength metasurfaces (devices composed of a metasurface patterned on both sides of a single substrate), and combined refractive-diffractive optics have been developed which utilize ray-tracing software such as Zemax Optics Studio [57], to determine the phase pattern required for the metasurface components. This approach has led to the development of wide field-of-view doublet metasurfaces [58,59], and achromatically corrected microscope objective lenses [60]. While this general workflow is appropriate for designing many useful optical components, it has the following shortcomings: Firstly, reliance on traditional ray-tracing software requires specifying the metasurface phase with smoothly varying basis functions, such as the Zernike polynomials. Potentially advantageous designs involving discontinuous phase functions are not considered. Secondly, polychromatic designs are challenging, since the achieved phase at each desired wavelength will be constrained by the nanoscale problem. One cannot independently solve the macroscale problem at each design wavelength, and then hope that a meta-atom geometry exists that achieves the correct phase at *all* desired wavelengths. Thirdly, while conventional lens design software is well suited for developing focusing optics and high-performance imaging systems, one may desire a metasurface to perform highly unusual functions where light-focusing may not be the intended effect—such as projecting unique intensity distributions at different wavelengths [61,62], or acting as a layer of an optical neural network [63,64]. Finally, for optical systems composed of many unique metasurface layers, we do not wish for the computational complexity of the underlying design problem to scale with the increasing number of design variables. We address each of these issues with our proposed inverse design method. Firstly, our method permits all design variables (the shapes of all meta-atoms within all metasurface layers of an optical system) to be computationally optimized independent of each other. Specification of a limited set of basis functions such as the Zernike polynomials is not required. Secondly, each iteration of our optimization method considers both the nanoscale and macroscale problem in tandem, so there is no risk of arriving at a solution to the macroscale problem, that cannot be realized due to the physical constraints of the nanoscale problem. This feature facilitates the design of polychromatic (and achromatic) devices. Thirdly, our design framework permits a user to specify an arbitrary desired output intensity distribution. Finally, as is the case with all inverse design methods, additional forward problem simulations are not required as the number of design variables increases. This feature permits thousands or millions of parameters to be optimized in parallel.

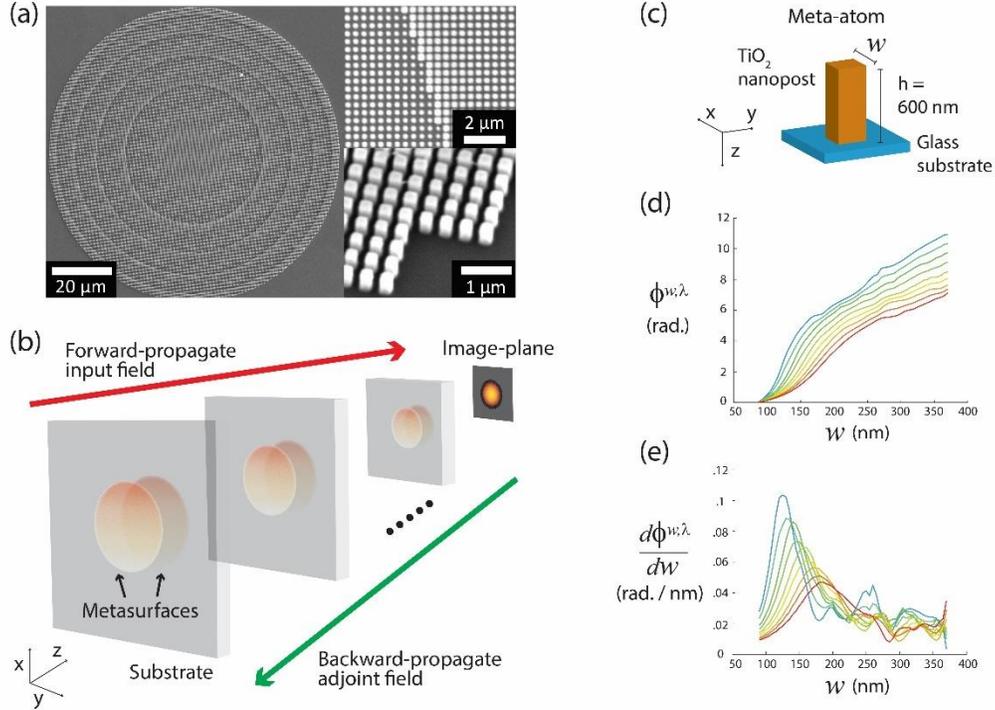

Fig. 1. Overview of metasurface optimization. (a) Electron micrograph of a representative metasurface lens, reprinted from [52]. This metasurface is composed of multiple $TiO_2$ nanopillars (meta-atoms) arranged in a rectangular lattice. By adjusting the widths of individual meta-atoms, the output phase is controlled. (b) General schematic of the types of designs considered in this paper: A series of glass substrates containing metasurfaces on either side are cascaded to form an optical system. Individual meta-atom geometries must be optimized to achieve a desired output intensity at the image plane. (c) In this paper, we consider square, $TiO_2$ meta-atoms of constant height, and tunable width $w$. Our simulations assume a lattice period of 400 nm. (d) An FDTD parameter sweep plots output phase as a function of meta-atom width. Results were smoothed using a moving average filter to remove resonances at specific widths/wavelengths. This dataset is used for all of the design examples included later in the paper. The individual plots are color-coded according to wavelength, with the bluest wavelength corresponding to $\lambda = 480$ nm, the reddest wavelength corresponding to $\lambda = 640$ nm, and 20 nm increments between wavelengths. (e) Using finite differences, the derivative of phase with respect to meta-atom width is estimated.

## 3. Mathematical framework

In this section, we develop the mathematical framework that underlies our approach to computational inverse design. We will first describe how a user of this framework can specify a cost function in terms of the squared errors between a desired set of output intensity distributions, and realized set of intensity distributions at a given design iteration. This cost function must then be iteratively minimized using the method of steepest-descent. In order to perform this optimization, it will be necessary to use a computationally efficient means of computing the gradient with respect to all of the design parameters (the individual meta-atom geometries). The gradient calculation is performed by solving a so-called adjoint problem within the context of a Fourier optics simulation and combining this result with an FDTD parameter sweep.

*3.1 Optimization Problem description*

Fig. 1(b) depicts the types of optical systems for which our design method is readily suited. We use a vector $\boldsymbol{E}^{in}$ to denote the electric field associated with light injected at an input plane of the optical system. Here, $\boldsymbol{E}^{in}$ is an $S \times 1$-dimensional (complex) vector containing discrete samples of the continuous input electric field distribution. In general, $\boldsymbol{E}^{in}$ may be a vectorization of the sampled 2D input plane for full 3D design problem, or 1D input plane for a 2D problem. The input field subsequently propagates through multiple layers of thin substrates coated with metasurfaces. At the plane of the *m*'th metasurface, the incident electric field accumulates a spatially-varying phase-change determined by the nanoscale geometric properties of the metasurface, which are denoted by the vector $\boldsymbol{w}^m$. Each metasurface consists of $S$ unique meta-atoms that can be independently parameterized by an arbitrary number of variables. For simplicity, here we only consider simple geometries (Fig. 1(c)), in which each meta-atom is a rectangular post of fixed height, and variable width (*w*). Hence, $\boldsymbol{w}^m$ will also be an $S \times 1$-dimensional (real) vector containing the widths of the individual meta-atoms. (Our method is applicable to arbitrarily complex meta-atom geometries, by increasing the dimensionality of $\boldsymbol{w}^m$ to store additional design parameters per individual meta-atom.) At the image plane of the optical system, the resulting electric field $\boldsymbol{E}^{out}$ will be a function of the parameters $\{\boldsymbol{w}^1, \boldsymbol{w}^2, \cdots, \boldsymbol{w}^M\}$ at each of the $M$ metasurfaces. The output intensity at the *s*'th sample position within the image plane is computed as $\boldsymbol{I}_s = \boldsymbol{E}_s^{out*} \boldsymbol{E}_s^{out}$. Subscripts denote individual elements of a given vector quantity. Our overarching goal will be to minimize the squared error between the realized intensity $\boldsymbol{I}$, and a user-defined *desired* intensity distribution $\boldsymbol{I}^{des}$ at each of the $S$ sample positions. Practically, the optical system should be designed to operate for different input light wavelengths $(\lambda)$ as well as different incident field distributions $(f)$ for each wavelength. Hence, the user may define multiple desired output intensity distributions as a function of wavelength and/or input fields. Given these specifications, we may formulate a cost-function $C$ that we hope to minimize, and an accompanying optimization problem:

$$\text{Minimize}: \quad C = \sum_{j=1}^{J} \sum_{k=1}^{K} \sum_{s=1}^{S} \left( \boldsymbol{I}_s^{\lambda_j, f_k} - \boldsymbol{I}_s^{des, \lambda_j, f_k} \right)^2 \quad (1)$$
$$\text{Subject to}: \quad w^{min} \prec \{\boldsymbol{w}^1, \boldsymbol{w}^2, \cdots, \boldsymbol{w}^M\} \prec w^{max}$$

That is, we minimize the summed squared error between the desired and realized output intensity at $J$ input wavelengths, $K$ input field distributions at a given wavelength, and $S$ sample points within the image plane. In our formulation of the optimization problem, we account for meta-atom geometric constraints $\{w^{min}, w^{max}\}$ which respectively specify the minimum and maximum width of a given meta-atom. In order to minimize $C$, one must iteratively adjust the design parameters $\{\boldsymbol{w}^1, \boldsymbol{w}^2, \cdots, \boldsymbol{w}^M\}$. It is desirable that this iterative optimization be performed using the steepest-descent algorithm [65]. At each step of the optimization, the design parameters are updated using the following rule:

$$\boldsymbol{w}^m = \boldsymbol{w}^m + \alpha \frac{dC}{d\boldsymbol{w}^m} \quad \forall m \quad (2)$$

Each vector of design parameters is adjusted by the gradient $dC/d\mathbf{w}^m$, scaled by a pre-defined step-size $\alpha$. The difficulty associated with this approach is the requirement that the gradient be computed efficiently, since each $\mathbf{w}^m$ may contain thousands or millions of design variables. Even for modestly-sized optical systems, brute force approaches are not practical. For example, an approximation of the gradient using finite-differences would entail independently adjusting each element of each $\mathbf{w}^m$ a small amount, and computing the resulting change in the cost function $C$. To estimate the gradient using this approach would entail a minimum of $M \times S + 1$ evaluations of $C$.

In the following sections, we will show how $dC/d\mathbf{w}^m$ may be computed in an efficient manner by solving an adjoint problem. This will enable steepest-descent optimization to be utilized for challenging design tasks. First, we will describe our forward problem, which is used to determine how light propagates through the optical system, and evaluate $C$ at a given design iteration. We will next derive an adjoint problem that has similar computational complexity to a single evaluation of $C$, but permits us to analytically evaluate the gradients with respect to all $\mathbf{w}^m$. Use of the adjoint-gradient method reduces the computational requirements of a single steepest-descent iteration from $M \times S + 1$ forward simulations (using finite differences) to just one forward simulation, and one adjoint simulation.

### 3.2 Forward problem formulation

Our forward propagation model is a discretized version of the angular spectrum wave propagator described in [66]. Briefly, the sampled electric field at the output of the *m*'th metasurface plane of the optical system is decomposed into a superposition of plane wave components using the discretized Fourier transform (this is practically accomplished using the Fast Fourier Transform Algorithm). Each plane wave (Fourier) component is multiplied by an appropriate phase factor that accounts for the phase accumulated as the wave propagates to the next surface within the optical system. An inverse Fourier transform operation is then performed to determine the resulting electric field. This propagation model permits large (~mm-scale) optical systems to be simulated efficiently, however a shortcoming of this approach is that multiple Fresnel-like reflections off of intermediate layers of the optical system are not modeled, leading to potential under-estimates of spurious background intensity at the image plane. Additionally, the Fourier optics framework does not capture near-field effects occurring at the metasurfaces, or between adjacent meta-atoms. At each metasurface plane, the incident electric field is multiplied by a diagonal matrix $\mathbf{\Phi}^{\mathbf{w}^m, \lambda_j}$ that contains complex exponentials associated with the phase delays induced by the choice of design variables $\mathbf{w}^m$ at input wavelength $\lambda_j$. We additionally find it helpful to define another diagonal matrix $\boldsymbol{\varphi}^{\mathbf{w}^m, \lambda_j}$ that contains the (real-valued) phase factors associated with $\mathbf{w}^m$. That is, $\mathbf{\Phi}^{\mathbf{w}^m, \lambda_j} = \exp\left(i\boldsymbol{\varphi}^{\mathbf{w}^m, \lambda_j}\right)$ for all diagonal entries of $\mathbf{\Phi}^{\mathbf{w}^m, \lambda_j}$, and zero otherwise. The components of $\mathbf{\Phi}^{\mathbf{w}^m, \lambda_j}$ and $\boldsymbol{\varphi}^{\mathbf{w}^m, \lambda_j}$ may be readily determined from a lookup table such as the one shown in Fig. 1(d). (For all of the design problems discussed in this paper, this lookup table was generated by performing a parameter sweep over titanium dioxide (TiO$_2$) nanopillars of different widths, a constant height of 600 nm, and a meta-atom pitch of 400 nm using the commercial software package Lumerical-FDTD [67], and measuring the output phase for all design wavelengths of interest.) We may express the forward propagation model for wavelength $\lambda_j$ and input field distribution $f_k$ as the following series of matrix-vector multiplications:

$$\boldsymbol{E}^{out,\lambda_j,f_k} = \left(\prod_{m=1}^{M} \boldsymbol{F}^{\dagger} \boldsymbol{P}^{m,\lambda_j} \boldsymbol{F} \boldsymbol{\Phi}^{w^m,\lambda_j}\right) \boldsymbol{E}^{in,\lambda_j,f_k} \tag{3}$$

In Equation (3), $\boldsymbol{F}$ and $\boldsymbol{F}^{\dagger}$ are the discrete Fourier transform matrix and its inverse ($\boldsymbol{F}$ is unitary, so its inverse is also its adjoint). $\boldsymbol{P}^{m,\lambda_j}$ is a diagonal matrix that effects the plane wave propagation of the individual Fourier components of the incident electric field. The $\{\xi,\xi\}$'th entry of this matrix is:

$$\boldsymbol{P}_{\xi,\xi}^{m,\lambda_j} = \exp\left(i\frac{2\pi z_m n_m}{\lambda_j}\sqrt{1-\left(\lambda_j \nu_\xi / n_m\right)^2}\right) \tag{4}$$

In Equation (4), $z_m$ is the axial distance from the $m$'th plane to the $(m+1)$'th plane in the optical system, and $n_m$ is the refractive index of the propagation medium (either air $n_m = 1$ or glass $n_m = 1.5$). $\nu_\xi$ is the spatial frequency corresponding to the $\xi$'th entry of the Fourier transformed electric field.

From Equation (3), it is straightforward to evaluate the output field $\boldsymbol{E}^{out,\lambda_j,f_k}$ from a given input field $\boldsymbol{E}^{in,\lambda_j,f_k}$. Once the output field is known, the output intensity and cost function $C$ may be evaluated. We note a couple differences between our formulation of the forward propagation model and other approaches [66]: First, we have chosen to use the angular spectrum wave propagator over the more familiar and computationally efficient Fresnel diffraction integral. The Fresnel diffraction integral requires only a single Fourier transform operation to propagate the electric field from one surface to the next, while our approach requires both a Fourier transform and inverse Fourier transform. However, Fresnel propagation assumes only paraxial wavefronts, while the angular spectrum wave propagator requires no such approximation (our approach is still a scalar-wave approximation and does not capture polarization effects). Secondly, previous work has utilized the Rayleigh-Sommerfeld diffraction integral to design multilayer optical systems in the context of training optical neural networks [63]. However, our approach enables the fast Fourier transform algorithm to be leveraged, amounting to significant computational savings-- $\mathrm{O}(MS\log(S))$ operations required to forward propagate the electric field, as opposed to $\mathrm{O}(MS^2)$ operations for an arbitrary series of $M$ matrix multiplications.

We point out some of the approximations that underlie our forward problem formulation: First, we assume that the output phase associated with a given meta-atom depends only on its own geometry, and is unaffected by its neighbors. We justify this approximation by reasoning that the optimal metasurface will be locally periodic over most regions. That is, the optimal metasurface geometry is usually not expected to change drastically as one moves a small number of lattice positions. Secondly, we assume uniform transmission for all metasurface geometries. That is, our optimization framework assumes that a change in meta-atom geometry will affect output phase, but not output intensity. Adjoint problem formations may be derived that consider both the effects of phase and intensity, but for simplicity and ease of mathematical calculations, we will restrict our optimization to consider only changes in phase. Once an optimized metasurface has been obtained, non-uniform transmission effects may be straightforwardly incorporated from FDTD simulations to calculate parameters such as focusing efficiency (see Section 3.1).

### 3.3 Adjoint problem derivation and fast calculation of gradients

Given a forward model for efficiently calculating $C$, we must now specify a method for computing the gradients $dC/dw^m$. We begin by dividing the gradient computation into a nanoscale problem that is computationally easy (when considering only simple metasurface structures), and a macroscale problem that is more computationally challenging. We may use the chain rule to express the gradient of $C$ as:

$$\frac{dC}{dw^m} = \frac{d\vec{\phi}^{w^m,\lambda_j}}{dw^m} \odot \frac{dC}{d\vec{\phi}^{w^m,\lambda_j}} \tag{5}$$

By some abuse of notation, we use the arrow annotation in $\vec{\phi}^{w^m,\lambda_j}$ to denote an $S\times 1$-dimensional vector containing the diagonal, non-zero entries of the matrix $\phi^{w^m,\lambda_j}$ (Hence, $\vec{\phi}^{w^m,\lambda_j}$ is a real-valued vector). The symbol $\odot$ denotes element-wise multiplication. The first (vector-valued) term on the right-hand side of Equation (5) is the nanoscale problem, and describes how a small change in the design variable $w^m$ will introduce a change in the phase at wavelength $\lambda_j$. This calculation may be easily performed by numerically differentiating the data shown in Fig. 1(d) using finite differences. The resulting numerical estimates are shown in Fig. 1(e). Determining how a local change in phase impacts the overall cost function $C$ (the second term in Equation (5)) is less straightforward. We will refer to the computation of $dC/d\vec{\phi}^{w^m,\lambda_j}$ as the macroscale problem, since the derivative with respect to the phase at any single metasurface position will depend upon all the other design variables.

We begin by considering the (scalar) derivative $dC/d\vec{\phi}_{s'}^{w^m,\lambda_j}$ at a single spatial location $s'$ on the $m$'th metasurface. We again make use of the chain rule:

$$\begin{aligned}\frac{dC}{d\vec{\phi}_{s'}^{w^m,\lambda_j}} &= \sum_{s=1}^{S}\left(\frac{dC}{dE_s^{out,\lambda_j,f_k}}\right)\left(\frac{dE_s^{out,\lambda_j,f_k}}{d\vec{\phi}_{s'}^{w^m,\lambda_j}}\right) \\ &= 4\Re\left\{\sum_{s=1}^{S}\left(E_s^{out,\lambda_j,f_k*}\left(I_s^{\lambda_j,f_k}-I_s^{des,\lambda_j,f_k}\right)\right)\left(\frac{dE_s^{out,\lambda_j,f_k}}{d\vec{\phi}_{s'}^{w^m,\lambda_j}}\right)\right\} \\ &= 4\Re\left\{\left(E^{out,\lambda_j,f_k}\odot\left(I^{\lambda_j,f_k}-I^{des,\lambda_j,f_k}\right)\right)^{\dagger}\left(\frac{dE^{out,\lambda_j,f_k}}{d\vec{\phi}_{s'}^{w^m,\lambda_j}}\right)\right\}\end{aligned} \tag{6}$$

In Equation (6), $\Re\{\ \}$ denotes the real portion of the terms enclosed in brackets. We must now use the forward problem formulation in Equation (3) to find a suitable expression for the derivative $dE^{out,\lambda_j,f_k}/d\vec{\phi}_{s'}^{w^m,\lambda_j}$:

$$\frac{dE^{out,\lambda_j,f_k}}{d\vec{\phi}_{s'}^{w^m,\lambda_j}} = \left(\prod_{m'=m}^{M} F^\dagger P^{m',\lambda_j} F \Phi^{w^{m'},\lambda_j}\right) \Delta^{s'} E^{m-1,\lambda_j,f_k} \tag{7}$$

Where $\Delta^{s'}$ is a diagonal matrix containing only one non-zero element. The $\{\xi,\xi\}$'th entry of this matrix is:

$$\Delta_{\xi,\xi}^{s'} = \begin{cases} i & \text{if } \xi = s' \\ 0 & \text{Otherwise} \end{cases} \tag{8}$$

This matrix appears in Equation (7) due to the fact that: $de^{ix}/dx = ie^{ix}$. The vector quantity $E^{m-1,\lambda_j,f_k}$ is the intermediate electric field incident upon the *m*'th metasurface of the optical system. That is:

$$E^{m-1,\lambda_j,f_k} = \begin{cases} E^{in,\lambda_j,f_k} & \text{if } m-1=0 \\ \left(\prod_{m'=1}^{m-1} F^\dagger P^{m',\lambda_j} F \Phi^{w^{m'},\lambda_j}\right) E^{in,\lambda_j,f_k} & \text{Otherwise} \end{cases} \tag{9}$$

Hence, by substituting Equation (7) into Equation (6), the derivative $dC/d\vec{\phi}_{s'}^{w^m,\lambda_j}$ may be evaluated:

$$\frac{dC}{d\vec{\phi}_{s'}^{w^m,\lambda_j}} = 4\Re\left\{\left(E^{out,\lambda_j,f_k} \odot \left(I^{\lambda_j,f_k} - I^{des,\lambda_j,f_k}\right)\right)^\dagger \left(\prod_{m'=m}^{M} F^\dagger P^{m',\lambda_j} F \Phi^{w^{m'},\lambda_j}\right) \Delta_{\xi,\xi}^{s'} E^{m-1,\lambda_j,f_k}\right\} \tag{10}$$

In general, this calculation would require a separate forward propagation through a portion of the optical system to evaluate just one of the $s'$ derivates contained in the gradient $dC/d\vec{\phi}^{w^m,\lambda_j}$. To improve computational efficiency, we now show how we may re-use intermediate results from a forward propagation of the field $E^{in,\lambda_j,f_k}$. Let us define an *adjoint field* as the the vector:

$$a^{\lambda_j,f_k} = E^{out,\lambda_j,f_k} \odot \left(I^{\lambda_j,f_k} - I^{des,\lambda_j,f_k}\right) \tag{11}$$

Since $dC/d\vec{\phi}_{s'}^{w^m,\lambda_j}$ is real, it is therefore equal to its complex conjugate. By taking the adjoint of Equation (10), and substituting the expression for the adjoint field (Equation (11)), we may write:

$$\frac{dC}{d\vec{\phi}_{s'}^{w^m,\lambda_j}} = 4\Re\left\{E^{m-1,\lambda_j,f_k\dagger} \Delta^{s'\dagger} \left(\prod_{m'=0}^{M-m} \Phi^{w^{M-m'},\lambda_j\dagger} F^\dagger P^{M-m',\lambda_j\dagger} F\right) a^{\lambda_j,f_k}\right\} \tag{12}$$

At first glance, evaluating Equation (12) has exactly the same computational complexity as evaluating Equation (10). However, this formulation may be leveraged as follows to more rapidly compute the *entire* gradient $dC/d\vec{\phi}^{w^m,\lambda_j}$: We recognize that the matrix $\Delta^{s'}$ effectively 'picks out' one element of a vector, and then multiplies that element by an additional factor of $i$. We therefore conclude:

$$\frac{dC}{d\vec{\phi}^{w^m,\lambda_j}} = -4\Re\left\{i\left(\boldsymbol{E}^{m-1,\lambda_j,f_k\dagger}\right)^T \odot \left(\prod_{m'=0}^{M-m}\Phi^{w^{M-m'},\lambda_j\dagger}\boldsymbol{F}^\dagger\boldsymbol{P}^{M-m',\lambda_j\dagger}\boldsymbol{F}\right)\boldsymbol{a}\right\} \quad (13)$$

Using Equations (5) and (13), the gradients at each metasurface plane may be calculated efficiently. The procedure is as follows: First, the input electric field $\boldsymbol{E}^{in,\lambda_j,f_k}$ is propagated through each metasurface layer of the optical system. At every metasurface plane, the intermediate electric field is stored for future use. Once the output field $\boldsymbol{E}^{out,\lambda_j,f_k}$ is determined, the adjoint field $\boldsymbol{a}^{\lambda_j,f_k}$ is then propagated *backwards* through the optical system. By performing element-wise multiplication of the intermediate (forward-propagated) fields with the (backward-propagated) adjoint fields at each metasurface plane, $dC/d\vec{\phi}^{w^m,\lambda_j}$ is evaluated. This quantity is then element-wise multiplied by $d\vec{\phi}^{w^m,\lambda_j}/dw^m$ to determine $dC/dw^m$. It is straightforward to appreciate the computational advantage of using this approach. Instead of performing a series of completely separate forward propagations to evaluate all of the $dC/d\vec{\phi}_{s'}^{w^m,\lambda_j}$, as would be required by Equation (10), we backward-propagate a single adjoint field, then perform a series of element-wise multiplications to determine the entries in the vector $dC/d\vec{\phi}^{w^m,\lambda_j}$ (Equation (13)). If we wished to design an optical system to operate at $J$ wavelengths and $K$ input fields at each unique wavelength, then a gradient calculation would require $J \times K$ forward propagations of the input fields, and $J \times K$ backward propagations (of similar computational complexity) of the adjoint fields. However, the required number of forward/backward propagations required would remain the same, regardless of the number of design variables used. In the following section, we will demonstrate the utility of this inverse design method, by using it to design optical devices for broadband imaging, spectral multiplexing, and optical neural networks.

**4. Computational inverse design examples**

In this section, we present three examples of novel optical systems designed using computational inverse design. All of the designs presented here utilize the same square $TiO_2$ on glass meta-atom library, and the FDTD results shown in Fig. 1(d). We first design an achromatic doublet metasurface lens or "metalens" that consists of two metasurfaces alternately patterned on the front and back side of a single glass substrate. Our design has a 160 nm bandwidth across the visible spectrum, an aperture diameter of 800 μm and a back focal length of 2 mm. The numerical aperture and diameter are superior to previous (experimentally demonstrated) achromatic devices. Furthermore, as our design does not utilize the geometric phase, or depend upon input light of a particular polarization state, focusing and transmission efficiency will be uniform for light of any input polarization. In our second example, we design a spectrally-multiplexed holographic device that produces a different output intensity distribution depending upon the wavelength of input light. We specify a five-color design for wavelengths each separated by 40 nm. Finally, due to the similarity between our computational inverse design method and the backpropagation algorithm used for neural networks, we use our approach to "train" an optical neural network composed of five glass substrates each patterned on either side with a unique metasurface. Our design may be used to classify handwritten digits projected onto the front of the device. Our proposed design can be trained in a computationally efficient manner, is suitable for visible wavelengths, and features a compact form-factor.

*4.1 Achromatic metalens doublet, with a large numerical aperture*

Chromatic aberration is a longstanding issue that impedes the adoption of metasurface optics for a wide range of imaging applications. Metasurface lenses designed for operation at a single wavelength usually focus red wavelengths a shorter distance than blue wavelengths – a highly undesirable feature for broadband imaging. (This phenomena is termed negative dispersion, and afflicts conventional diffractive elements in addition to metasurfaces [68-70].) In the context of metasurfaces, current approaches to correcting chromatic aberration involve spatial segmentation [71], layered polychromatic devices [72], use of the Pancharatnam-Berry (geometric) phase [73-76], polarization rotation [52,77], dispersion engineering with reflective substrates [78,79], and computational deconvolution combined with extended-depth-of-field metasurfaces [80]. Related computational design methods for broadband diffractive optical elements have also been proposed [81]. Previously reported metasurface achromats have small diameters, low numerical apertures, may require a specific input polarization and often suffer from low overall transmission efficiency. Here, we present a design for an achromatic metasurface doublet lens that is polarization independent. Furthermore, our design uses simple rectangular nanostructures, which would potentially make fabrication easier.

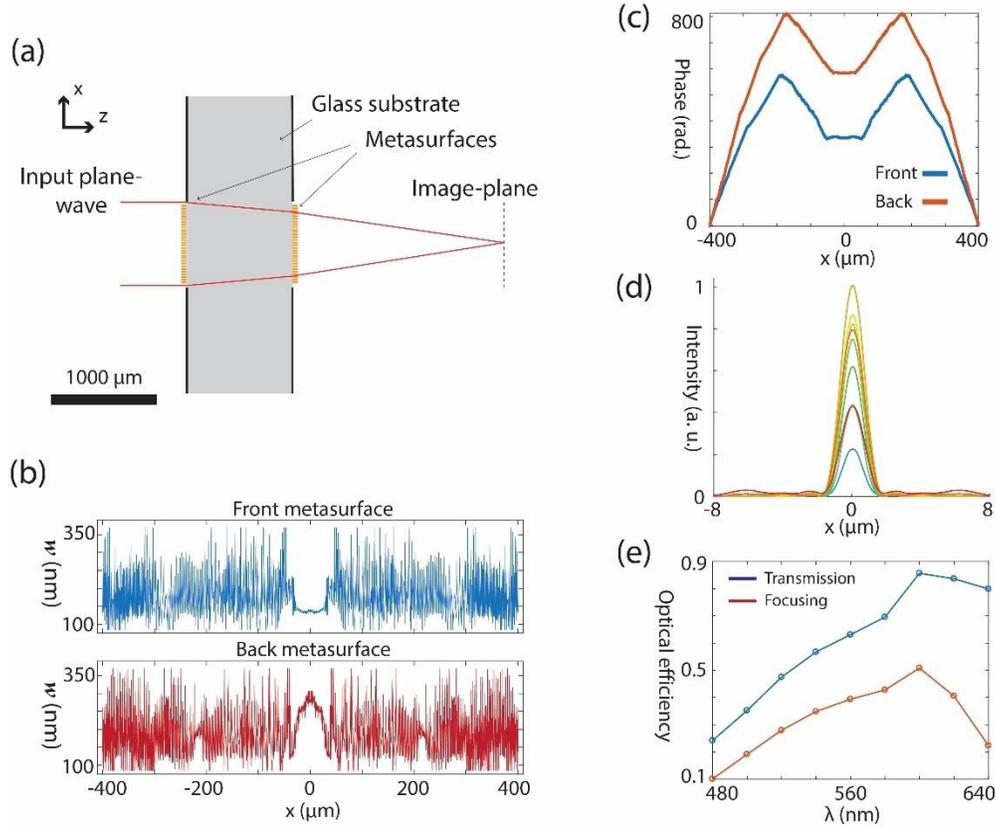

Fig. 2. Inverse design of an achromatic doublet. (a) Axial schematic of design. (b) Optimized meta-atom widths of front and back metasurface. Widths of individual meta-atoms are plotted as a function of aperture position. (c) Unwrapped phase across metasurface apertures at $\lambda = 640$ nm. (d) Cross-section of focal plane intensity at all design wavelengths. Individual plots are color-coded according to wavelength, with the bluest wavelength corresponding to $\lambda = 480$ nm, the reddest wavelength corresponding to $\lambda = 640$ nm, and 20 nm increments between wavelengths. (e) Transmission and focusing efficiency calculations at all design wavelengths.

A schematic of our design is shown in Fig. 2(a). Both sides of a 1 mm thick glass substrate are patterned with a unique metasurface. It is desired that normally-incident broadband light come to a focus 2 mm beyond the second metasurface (the backside of the glass substrate). We specify that the metasurface aperture be 800 µm in diameter. If we perform optimization across a single line through the center of the optic (i. e. we explicitly perform optimization within a two-dimensional (XZ) simulation region) and assume that the metalens is symmetric, it follows that we must optimize two thousand design parameters (given a 400 nm pitch, one thousand unique meta-atom widths must be determined for both the front and back metasurface of the optic). To specify a target intensity distribution for our design, $I^{des,\lambda_j}$, we calculated the ideal, diffraction-limited focal spot associated with a singlet lens with focal length 2 mm, and aperture diameter 800 µm, (the desired intensity distribution was changed for each input wavelength). For our input fields, we chose normally-incident planewaves with wavelengths λ = 480 nm through 640 nm in 20 nm increments (nine input fields in total). Forward propagation and adjoint calculations were performed in a two-dimensional simulation (e. g. along a central cross-section of the metalens). To perform steepest-descent iterations, a separate adjoint simulation was solved for each of the nine design wavelengths. The gradients calculated at each wavelength were summed to determine the overall gradient $dC/d\bm{w}^m$. All meta-atom widths were initially set to 200 nm, (ensuring an initially flat phase response for all design wavelengths). Optimization proceeded by rescaling the gradients such that the maximum change in the width of any meta-atom was set to 5 nm per iteration, and 100 iterations were performed in this manner. The maximum width change was then set to 1 nm, and 500 more iterations were performed. To enforce manufacturing constraints, we specified that the minimum meta-atom width was $w^{min}$ = 85 nm, and the maximum was $w^{max}$ = 370 nm. If a given steepest-descent iteration caused a meta-atom width to move outside of these constraints, the step was truncated such that that meta-atom was instead set to either $w^{min}$ or $w^{max}$.

The resulting optimized meta-atom widths for the front and back metasurfaces are plotted in Fig. 2(b). Further intuition about the resulting design can be gained by plotting the unwrapped phase across the metasurface aperture at a single wavelength (λ=640 nm, see Fig. 2(c)). Unlike conventional lenses that utilize a parabolic or spherical phase to focus light, Our design features sharp changes in the gradient of the phase at a radial distance ~180 µm from the center of the metasurface aperture. Such sudden phase discontinuities would be difficult to realize using e. g. Zernike polynomials, and conventional optical design software—highlighting a key advantage of an inverse design approach that independently optimizes all design variables. To estimate the overall optical efficiency of our design after optimization, transmission data (determined from the FDTD parameter sweep) for each optimized meta-atom width was incorporated into our Fourier optics simulations by multiplying the incident electric field at each metasurface layer with a diagonal matrix containing the spatially-varying transmission amplitudes. In Fig. 2(d), the simulated intensity across the center of the image plane is plotted for each of the design wavelengths. We achieve nearly diffraction-limited focusing performance across a relatively wide bandwidth of 160 nm. In Fig. 2(e), overall optical transmission through the two metasurface layers is plotted as a function of wavelength. In addition, we plot focusing efficiency, defined as the fraction of incident intensity contained in a diffraction-limited region (~1.6 µm radius) at the image plane. Overall, focusing efficiency is reduced in comparison to singlet metasurfaces designed for single-wavelength operation [5]. However, performance is superior to many existing achromatic designs over comparable bandwidths. Furthermore, recent work [30] has demonstrated that high focusing efficiency may be achieved by optimizing the geometries of individual meta-atoms (of sizes ~2.5λ) using adjoint-gradient methods. Hence, we speculate that focusing efficiency could be further enhanced by combining such nanoscale optimization methods with our approach. XZ-profiles of the optimized metalens point spread function are shown in Fig. 3(a). These plots confirm that for each of the design wavelengths, peak intensity is directed at a plane 2 mm from the rear surface of the metalens. To highlight how the interplay between the front and back metasurface is necessary for achieving broadband performance, we re-ran the optimization procedure, but optimized only with respect to the rear metasurface (uniform phase and perfect transmission were assumed at the surface that would normally contain the front metasurface). XZ-profiles of the point spread function associated with the singlet design are shown in Fig. 3(b). Intensity distributions are plotted on the same intensity scale as in Fig. 3(a). For the case of optimizing a metasurface singlet lens, focal plane intensity is severely reduced, and spurious intensity peaks occur at multiple axial positions removed from the focal plane. These results demonstrate how systems composed of multiple metasurfaces introduce an expanded range of capabilities.

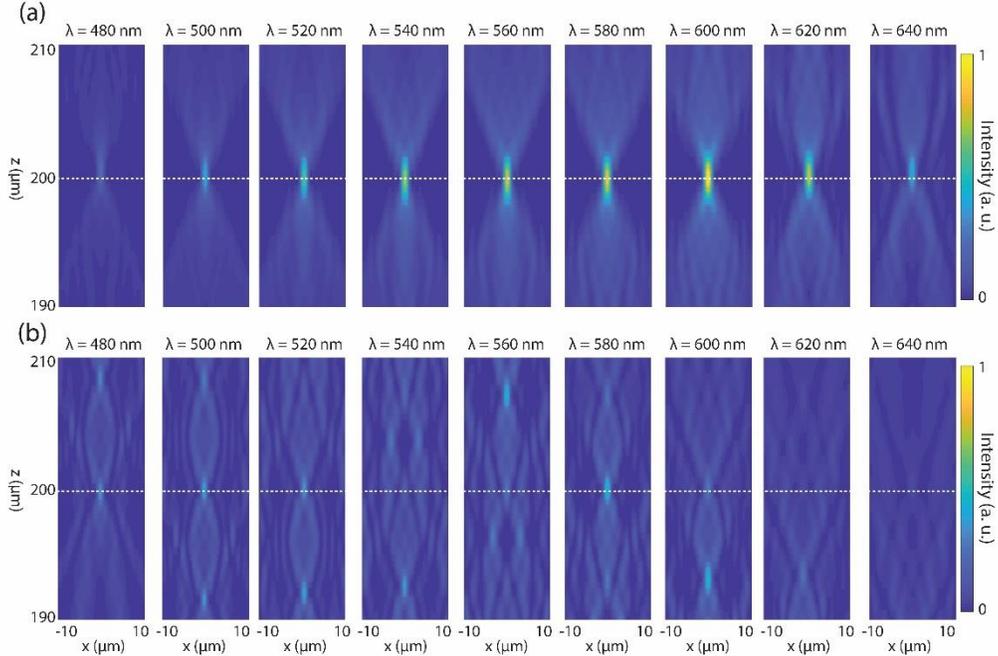

Fig. 3. XZ-slices of the point spread function for doublet and singlet metalenses. (a) XZ-slices of focal-region intensity for an optimized metasurface doublet. (b) XZ-slices of focal-region intensity for an optimized metasurface singlet. Intensity color scale for (a) and (b) are identical.

*4.2 Generation of arbitrary, spectrally-multiplexed intensity distributions*

A key advantage of the inverse design method is that it permits a user to define an arbitrary set of output intensity distributions at the image plane $\left(\boldsymbol{I}^{des,\lambda_j,f_k}\right)$. There is no need to restrict the desired output to the functions performed by conventional optical elements. To illustrate this feature, we show how a single metasurface device may be used to generate a set of unique output intensities, depending solely upon the input wavelength. In the context of silicon photonics, devices with wavelength-splitting functionality have been designed using electromagnetic inverse solvers [22,25]. Furthermore, wavelength-multiplexed metasurfaces could find application in fields such as fluorescence bioimaging, where under some circumstances it is advantageous to encode wavelength-specific optical aberrations into acquired images for classification and 3D localization tasks [82-85].

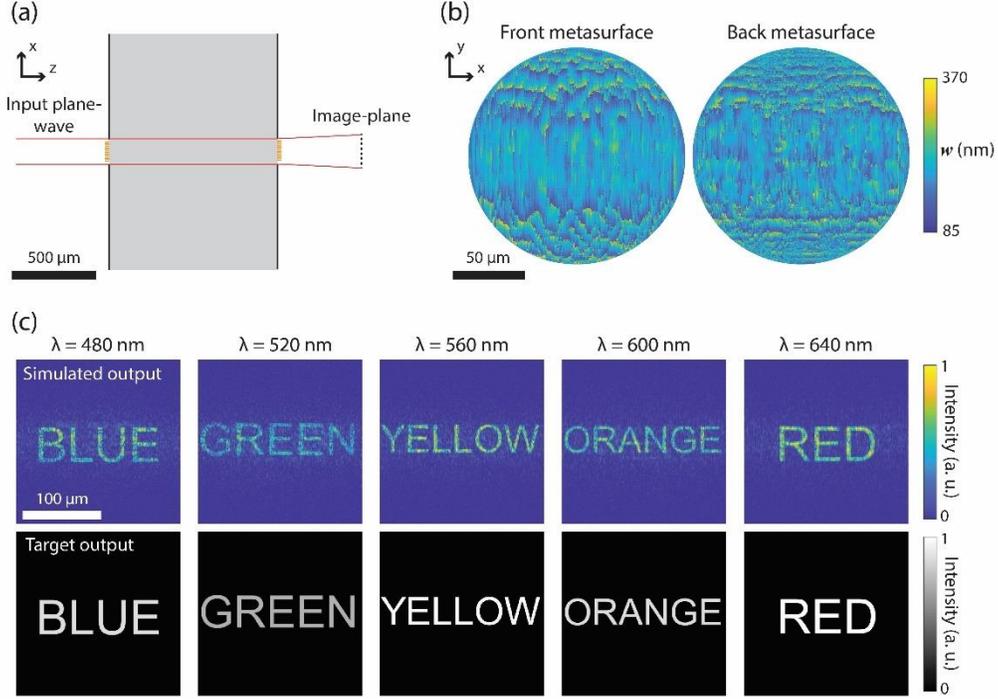

Fig. 3. Inverse design of a spectrally-multiplexed holographic element. (a) Axial schematic of design. (b) Optimized design of front and back metasurfaces. Widths of individual meta-atoms are plotted. (c) Top: Simulated output intensity distributions for optimized design. Bottom: Target output intensity distributions.

A schematic of our design is illustrated in Fig. 4(a). Input light (assumed to be spatially coherent) is normally incident on a 1 mm thick glass substrate coated with a metasurface on the front and back face. Here we perform optimization over a full three-dimensional simulation region. Each of the two metasurfaces is 150 μm in diameter and has a circular aperture within an XY plane. A wavelength-dependent desired image is expected to form 500 μm beyond the back metasurface. We again assume a 400 nm pitch between adjacent meta-atoms. Hence, each meta-surface contains 110,461 meta-atoms, and the optimization problem will consist of a total of 220,922 independent design variables. We specify five unique input wavelengths (λ = 480 through 640 nm in 40 nm increments), and five unique desired output intensity distributions. Optimization consisted of 500 steepest descent iterations with a maximum change in meta-atom width of 5 nm per iteration. Unlike in the previous example, in this case it was necessary to solve forward and adjoint problems over a full three-dimensional simulation region.

The optimized meta-atom widths (Fig. 4(b)) show that the optimized design makes use of all degrees of freedom offered by the numerous design variables. Output intensity distributions for the five input wavelengths are plotted in Fig. 4(c). The output intensities spell the words BLUE, GREEN, YELLOW, ORANGE and RED, for different input wavelengths. To estimate the overall optical efficiency $O^{\lambda_i}$ of this holographic element, for each of the five output images the following sum based on the root-mean square error was calculated:

$$O^{\lambda_i} = 1 - \sqrt{\frac{\sum_{s=1}^{S}\left(\bm{I}_s^{\lambda_i} - \bm{I}_s^{des,\lambda_i}\right)^2}{\sum_{s=1}^{S}\left(\bm{I}_s^{des,\lambda_i}\right)^2}} \qquad (14)$$

For the wavelengths 480 nm, 520 nm, 560 nm, 600 nm, and 640 nm, $O^{\lambda_i}$ was calculated as 40.6%, 37.4%, 45.6%, 43.5%, and 47.7% respectively. It is important to note that the exact user-specified target intensity distributions (bottom, Fig. 4(c)) are physically infeasible, due to the fact that the target distribution contains discontinuous transitions between zero and non-zero intensity over a subwavelength sampled grid. Nevertheless, the design method is able to handle this physically-unrealizable input gracefully and return a result that matches the target intensity as closely as possible.

*4.3 Design of an optical neural network for image classification*

As a final demonstration of our inverse design method, we show how it may be used to train an optical neural network to classify handwritten digits 0 through 9. Optical neural networks are an emerging field that has recently generated much interest within the machine learning and artificial intelligence communities [63,64,86,87]. Conventional neural networks are implemented entirely in software and perform a series of complex mathematical operations on input data in order to assign the data to one of many possible output classes. Alternatively, optical neural networks utilize light propagation to perform a multitude of mathematical operations and use the intensity distribution at the output of the network to perform classification tasks, or as an intermediate input for further computation performed in software. Optical neural networks effectively perform computations such as Fourier transforms and convolutions "at the speed of light". Hence, an advantage to implementing a neural-network partially or completely by optical means is the potential to drastically speed up and further parallelize computations on large datasets. Previous work has experimentally demonstrated how image classification tasks may be performed optically [63,64]. Neural network training may also be done by optical means [87]. Here, we recognize that our inverse-design method operates in a manner similar to the conventional backpropagation algorithm used for training (software) neural networks. However, our method further increases training speed. In a standard linear network, forward propagation and gradient calculation requires $\mathrm{O}(MS^2)$ operations, where $S$ is the number of meta-atoms (neurons) contained in a single metasurface (layer) of the network, and $M$ is the number of layers. However, since our method uses the fast Fourier transform algorithm to simulate propagation of input and adjoint fields, computation time is reduced to $\mathrm{O}(MS\log(S))$ operations.

Our proposed design is shown schematically in Fig. 5(a). Pixilated images of handwritten digits are projected upon the input aperture of the device using monochromatic, coherent light (λ = 560 nm). The neural network consists of five glass substrates 100 μm thick, axially arranged with 100 μm air separations. A 150 μm diameeter circular metasurface is patterned on the front and back of each element, such that the entire neural network contains ten unique layers. After light exits the final glass substrate, it propagates 0.5 mm, until it impinges upon an array of ten detectors, arranged in a circular configuration at the image plane. Each detector is assigned a specific digit. Image classification proceeds by measuring the total intensity incident upon each of the ten detectors, and assigning input images to the digit class that has the greatest associated output intensity (note that this classification strategy is effectively the same as that used by [63]).

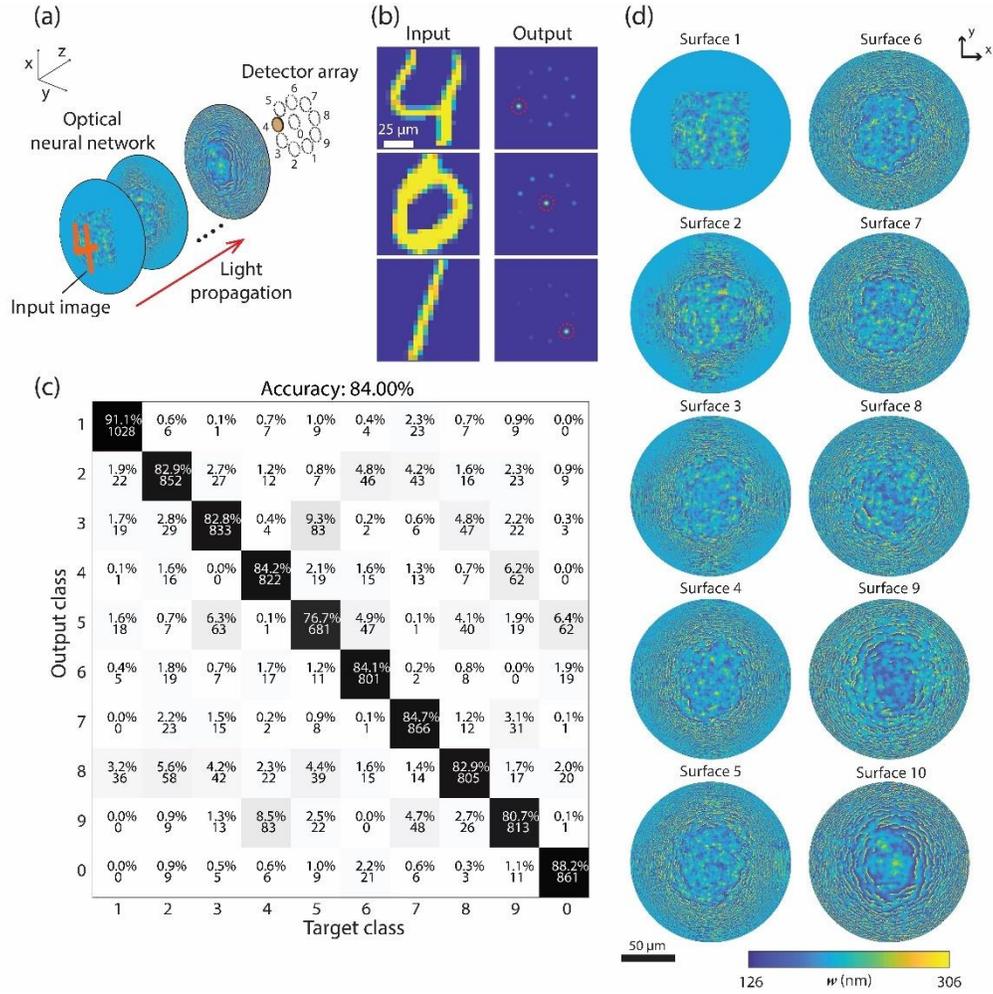

Fig. 5. An optical neural network for handwritten digit classification. (a) Schematic of design. Handwritten digits are projected on the front of the network. Light propagates through the optical system and is incident on an array of ten detectors at the image plane. Digits are classified based on which detector receives the most intensity. (b) Simulated test inputs and outputs of optimized design. A different detector receives the majority of output intensity, depending upon which digit is projected at the input. (c) Confusion matrix for MNIST testing dataset. (d) Optimized metasurface designs. Widths of individual meta-atoms are plotted.

Given this optical system layout, the next task is to train the neural network by optimizing individual meta-atom dimensions such that different input intensity distributions illuminate the correct output detectors. This optimization problem involves 1,104,610 design variables (110,461 meta-atoms per metasurface, and ten metasurfaces). Training was performed using the 60,000 handwritten digit images contained in the MNIST training dataset [88]. Digits were rescaled such that they encompassed a bounding box of 75 μm diameter. It is assumed that each of the digits contained in the training dataset was appropriately centered within the bounding box. For each training image $k$, an output intensity distribution $\left(\boldsymbol{I}^{des,f_k}\right)$ was defined as a two-dimensional Gaussian centered over the appropriate output detector with variance parameter $\sigma^2 = 2.25$ μm$^2$. Optimization of the design proceeded using the stochastic steepest-descent method. The 60,000 training images and associated desired outputs were divided into batches of five images, and training consisted of five epochs. For the first epoch, a maximum step-size (maximum change of meta-atom width per steepest descent iteration) of 5 nm was used. For the second epoch a 1 nm step size was used, and for the final three epochs a 0.1 nm step size was used. Fig. 4(d) plots the widths of the meta-atoms of the final design. To gain an intuitive understanding of the overall function of this optical system, Fig. 5(b) plots three representative input digits, and the resulting simulated output intensity distributions. For all digits, the majority of the output intensity is concentrated in ten regions centered over the ring of detectors. However, in each of the cases shown, a different detector receives the greatest overall intensity, enabling each of the unique digits to be classified correctly.

Once an optimized design was arrived at, the performance of the system was evaluated by attempting to classify the 10,000 handwritten digits contained in MNIST testing dataset. On the test data, we achieve a classification accuracy of 84.00%. A confusion matrix for our result is plotted in Fig. 5(c). The optical network has greatest difficulty distinguishing between "4" and "9", and "5" and "3", based on the similarities in the shapes of these digits. In comparison, [63] obtained a simulated test accuracy of 91.75%. This difference in performance could be due to factors such as the overall geometry and dimensions of our design, or differences in the initialization and training routine. In closing this section, we note that even though our optical design consists of multiple layers of metasurfaces, the neural network is effectively still a linear system [89]. That is, each metasurface layer and subsequent propagation can be thought of a linear transformation upon the incident intensity distribution. Use of optical nonlinearities [90-92] could potentially improve overall performance.

## 5. Discussion

We have developed an inverse design method for optimizing cascaded systems of metasurface optics. The method is computationally efficient and realizes a variety of innovative design solutions. Each of the designs presented here utilizes a $TiO_2$-on-glass fabrication platform based on square meta-atoms arranged in a periodic lattice. Hence the only adjustable variable for each individual meta-atom is its associated width. This simple fabrication platform has a number of advantages: The manufacturing feasibility of the meta-atom geometries is well established, the four-fold symmetry of the meta-atoms ensures polarization-independent performance, and the total number of design variables remains relatively modest. Nevertheless, it is reasonable to expect that performance of the example designs considered could be further enhanced if more complex meta-atom geometries were considered throughout the design optimization routine, especially in the context of polychromatic and achromatic designs. For example, multi-layer (singlet) metasurfaces, and meta-atom libraries of crosses, hollow rectangles and concentric cylinders have been investigated [56]. We briefly discuss a couple ways in which our inverse design framework could be extended to include this additional functionality. The most straightforward approach to consider more complex meta-atom geometries would be to perform a multi-dimensional FDTD parameter sweep over the multiple design variables. In this case, estimating the gradients of the nanoscale problem (calculation of $d\varphi^{w^m, \lambda_j}/dw^m$, see Equation (5)) would require computing finite differences over a multi-dimensional parameter space as opposed to a single width parameter, but the overall method would remain nearly unchanged. An obvious drawback to this strategy is that a complete parameter sweep becomes computationally intractable as the number of design variables per meta-atom increases. In the FDTD parameter sweeps that we have performed for this paper, a single meta-atom geometry can take ~minutes to simulate using Lumerical FDTD on most standard desktop workstations. Hence, a well-sampled sweep over three or more design variables could easily take ~weeks to complete, without utilizing high-performance computing resources. An approach that would not require an exhaustive parameter sweep would be to solve a set of different adjoint problems at each design iteration: First, a macroscale adjoint problem using the framework described here would be solved to determine appropriate gradients with respect to phase at each meta-atom location within the optical system. Initial meta-atom geometries could then be selected from a (relatively small) library of simple shapes such as rectangles or cylinders. Next, using some suitable error threshold, lattice positions within the metasurface are identified where varying the parameters of the simple meta-atom structures does not produce desired gradients with respect to phase. At these locations, a separate set of nanoscale adjoint problems could be solved to determine gradients in phase with respect to a larger set of design parameters, and meta-atom shapes updated accordingly. As each nanoscale adjoint problem would be performed over the dimensions of a single meta-atom, this process would not require unreasonable computational resources, and could be trivially parallelized. Means of formulating appropriate nanoscale adjoint problems have been described in previous literature [19-21], and commercial software packages for implementing adjoint solvers in FDTD are currently available [67]. Our inverse design framework, combined with the possible extensions discussed here may lead to numerous applications for metasurfaces to the fields of imaging, display, and optical computing.

## Funding

The author gratefully acknowledges research funding from the Harry S. Truman Fellowship at Sandia National Laboratories. This work was supported by the Laboratory Directed Research and Development program at Sandia National Laboratories, a multimission laboratory managed and operated by the National Technology and Engineering Solutions of Sandia, LLC, a wholly owned subsidiary of Honeywell International Inc., for the U.S. Department of Energy's National Nuclear Security Administration under contract no. DE-NA0003525. This paper describes objective technical results and analysis. Any subjective views or opinions that might be expressed in the paper do not necessarily represent the views of the U.S. Department of Energy or the U.S. government.

## Acknowledgements

The author wishes to acknowledge Michael B. Sinclair for discussions regarding the adjoint problem derivation and careful reading of the manuscript. The author thanks Maxwell D. Aiello for providing the representative SEM images of a metasurface shown in Fig. 1(a) of the text. The author also acknowledges Victor M. Acosta, John D. Perreault, and Patrick Llull for helpful conversations regarding computational inverse design and its applications.

74. S. Wang, P. C. Wu, V. Su, Y. Lai, M. Chen, H. Y. Kuo, B. H. Chen, Y. H. Chen, T. Huang, J. Wang, R. Lin, C. Kuan, T. Li, Z. Wang, S. Zhu, and D. P. Tsai, "A broadband achromatic metalens in the visible," Nat. Nanotech. **13**, 227-232 (2018).
75. W. T. Chen, A. Y. Zhu, V. Sanjeev, M. Khorasaninejad, Z. Shi, E. Lee, and F. Capasso, "A broadband achromatic metalens for focusing and imaging in the visible," Nat. Nanotech. **13**, 220-226 (2018).
76. W. T. Chen, A. Y. Zhu, J. Sisler, Z. Bharwani, and F. Capasso, "A broadband achromatic polarization-insensitive metalens consisting of anisotropic nanostructures," Nat. Commun. **10**, 355 (2019).
77. E. Arbabi, A. Arbabi, S. M. Kamali, Y. Horie, and A. Faraon, "High efficiency double-wavelength dielectric metasurface lenses with dichroic birefringent meta-atoms," Opt. Express **24**(16), 18468-18477 (2016).
78. E. Arbabi, A. Arbabi, S. M. Kamali, Y. Horie, and A. Faraon, "Controlling the sign of chromatic dispersion in diffractive optics with dielectric metasurfaces," Optica **4**(6), 625-632 (2017).
79. M. Khorasaninejad, Z. Shi, A. Y. Zhu, W. T. Chen, V. Sanjeev, A. Zaidi, and F. Capasso, "Achromatic metalens over 60 nm bandwidth in the visible and metalens with reverse chromatic dispersion," Nano Lett. **17**(3), 1819-1824 (2017).
80. S. Colburn, A. Zhan, and A. Majumdar, "Metasurface optics for full-color computational imaging," Sci. Adv. **4**(2), eaar2114 (2018).
81. P. Wang, N. Mohammad, and R. Menon, "Chromatic-aberration-corrected diffractive lenses for ultra-broadband focusing," Sci. Rep. **6**, 21545 (2016).
82. C. Smith, M. Huisman, M. Siemons, D. Grünwald, S. Stallinga, "Simultaneous measurement of emission color and 3D position of single molecules," Opt. Express **24**(5), 4996-5013 (2016).
83. A.Jesacher, S. Bernet, and M. Ritsche-Marte, "Colored point spread function engineering for parallel confocal microscopy," Opt. Express **24**(24), 27395-27402 (2016).
84. Y. Shechtman, L. E. Weiss, A. S. Backer, M. Y. Lee, and W. E. Moerner, "Multicolor localization microscopy by point-spread-function engineering," Nat. Photonics **10**, 590-594 (2016).
85. E. Hershko, L. E. Weiss, T. Michaeli, and Y. Shechtman, "Multicolor localization microscopy and point-spread-function engineering by deep learning,"Opt. Express **27**(5), 6158-6183 (2019).
86. Y. Shen, N. C. Harris, S. Skirlo, M. Prabhu, T. Baehr-Jones, M. Hochberg, X. Sun, S. Zhao, H. Larochelle, D. Englund, and M. Soljačić, "Deep learning with coherent nanophotonic circuits," Nat. Photonics **11**, 441-446 (2017).
87. T. W. Hughes, M. Minkov, Y. Shi, S. Fan, "Training of photonic neural networks through *in situ* backpropagation and gradient measurement," Optica **5**(7), 864-871 (2018).
88. Y. LeCun, L. Bottou, Y. Bengio, and P. Haffner, "Gradient-based learning applied to document recognition," Proc. IEEE **86**(11), 2278-2324 (1998).
89. H. Wang, G. Huang, X. Wei, Y. Sun, and H. Wang, "Comment on 'all-optical machine learning using diffractive deep neural networks,'" https://arxiv.org/abs/1809.08360 (2018).
90. S. Liu, P. P. Vabishchevich, A. Vaskin, J. L. Reno, G. M. Peake, G. A. Keeler, M. B. Sinclair, I. Staude, I. Brener, "Optical nonlinearities in all-dielectric metasurfaces," 2018 International Conference on Optical MEMS and Nanophotonics doi:10.1109/OMN.2018.8454615.
91. M. Miscuglio, A. Mehrabian, Z. Hu, S. I. Azzam, J. George, A. V. Kildishev, M. Pelton, V. J. Sorger, "All-optical nonlinear activation function for photonic neural networks," Opt. Mater. Express **8**(12), 3851-3863 (2018).
92. Y. Zuo, B. Li, Y. Zhao, Y. Jiang, Y. Chen, P. Chen, G. Jo, J. Liu, S. Du, "All optical neural network with nonlinear activation functions," https://arxiv.org/abs/1904.10819 (2019).
21